\documentstyle [aps,manuscript,psfig,epsfig]{revtex}

\def\p{\mbox{$\partial$}}
\newcommand{\be}    {\begin{equation}}
\newcommand{\ee}    {\end{equation}}

\begin{document}

\title{Two-dimensional Navier--Stokes simulation of breaking waves}

\author{Gang Chen and Christian Kharif}

\address{Institut de Recherche sur les Ph\'enom\`enes Hors Equilibre,
 UMR CNRS 138, Case 903, \\
163, avenue de Luminy,
13288 Marseille Cedex 9, France}

\author{St\'ephane Zaleski and Jie Li}

\address{Laboratoire de Mod\'elisation en M\'ecanique, URA CNRS 229,
Universit\'e Pierre et Marie Curie,\\ 4 place Jussieu,
75252 Paris Cedex 05, France}

\maketitle

\begin{abstract}
 Numerical simulations describing plunging breakers including the
splash-up phenomenon are presented. The motion is governed by the classical, 
incompressible, two-dimensional Navier-Stokes equation. The numerical
 modelling of this two-phase flow is based on a piecewise linear version of 
the volume of fluid method. Capillary effects are taken into account as a 
stress tensor computed from gradients of the volume fraction function.
 Preliminary results  concerning the time evolution of liquid--gas 
interface and vorticity field are given for short waves, showing how an
 initial steep wave undergoes breaking and successive splash-up cycles. 
Different evolutions of the wave  energy are observed  during the breaking 
stage. The energy dissipation due to viscosity becomes significantly 
important at each time of the impact of jet and the formation of splash-up.
 It is found that nearly $70\%$
of the wave energy is lost after about three periods.
\end{abstract}

\newpage

Plunging breakers are due to the formation of a jet at the crest of the wave
or  in its vicinity. Under the influence of gravity the jet plunges down
into the water generating a splash-up phenomenon and important turbulence 
generation (see for example Bonmarin${}^1$). Generation of bubbles 
and spray is observed.
Numerical experiments describing the evolution of breaking waves up to the
 time of impact of the jet have been developed successfully 
by numerous investigators using methods based on potential flows.  Among the 
main  contributions are the works of Longuet-Higgins and 
Cokelet,${}^2$ Vinje and Brevig,${}^3$
Baker {\it et al.},${}^4$ and New {\it et al.}.${}^5$ This list is not 
exhaustive (see for more detailed reviews Peregrine,${}^6$ and Banner and 
Peregrine${}^7$). The process of the initiation of spilling
 breakers is less well understood. 
Two main mechanisms have been proposed. Computations of overturning waves
by  New {\it et al.}${}^5$ demonstrated that there is no essential difference
between spilling and  plunging breakers:  the only difference is  scale
of the jet at the wave crest. This led these authors to suggest
that spilling breakers might be produced by a small plunging event at
the wave crest. More recently, Longuet-Higgins and Cleaver${}^8$
pointed out that a ``crest instability'' may represent the initial stage of the
spilling breaker which has been found in laboratory by 
Duncan {\it et al.}.${}^9$ This instability causes a bulge to form
ahead the crest with capillary waves just below the ``toe''. 

There is little numerical simulation of this problem going beyond the
time of impact of the plunging jet. In Monaghan {\it et al.},${}^{10}$ the
smoothed particle hydrodynamics method (SPH) was used to simulate
a splash-up. However the spatial resolution of the method seems insufficient
to resolve the small scale viscous and capillary effects.
The aim of the present work is to attempt a preliminary numerical
solution of the Navier--Stokes equations of this problem, on grids
 sufficiently fine so that viscous and
capillary effects could be retained. We wish to simulate
the two-phase flow of bubbles and droplets following splash-up
as well as vortex-like motions. To this effect we have used a robust
 numerical technique, based on the volume of fluid method${}^{11-16}$
 that allows us to use rather large grids and to follow large interface 
deformations and topology changes.

We consider viscous incompressible flow with a constant surface tension 
$\sigma$ at the liquid-gas interface. The governing equations for the velocity
vector ${\bf u}=(u,v)$ and pressure $p$ in the bulk of each phase
 (liquid and gas) are the classical Navier-Stokes equations supplemented 
by the condition of incompressibility:
\be
\rho \left( \p_t {\bf u} +  {\bf u} \cdot \nabla \,{\bf u}\right ) =
 -\nabla p + \nabla \cdot (\mu{\bf S}) + \rho{\bf g} +
 \sigma\kappa\delta_S{\bf n}, \label{N-S}
\ee
\be
\nabla \cdot {\bf u} = 0 . \label{incom}
\ee
Here $\rho$ is the density, $\mu$ is the dynamic viscosity, ${\bf S}$ is
the rate of strain tensor, whose components are $S_{ij}=\p u_i/\p x_j +
\p u_j /\p x_i$, $\kappa$ is the curvature,
and ${\bf n}$ is the a normal to the interface. $\delta_S$ is a delta function 
that is zero everywhere except at the interface. ${\bf g}=(0,-g)$ is the
acceleration due to gravity. The location of the two fluids is specified 
with the help of a volume fraction function $C$, with $C=1$ inside one fluid
 (liquid for example) and $C=0$ in the other. An interface is to be
 constructed only if the fraction $C$ is between $0$ and $1$. This is made 
by a volume of fluid type numerical technique, which is now well
 documented.${}^{11-16}$
This method allows us to follow interfaces beyond the point of reconnection,
and is relatively simple and robust. 

To simulate the wave kinematics and dynamics in infinite depth, we assume
the motion to be two-dimensional and  spatially periodic.
 The corresponding computational domain is a
square cavity of length  equal to the wavelength, $\lambda$,
with periodic boundary condition in the direction of wave propagation 
and free-slip condition along the others. The reference length and time
are chosen as $\lambda$ and $(\lambda/g)^{1/2}$, respectively. As a result,
four dimensionless parameters appear in the governing equations. They are
the Reynolds number, $Re=\rho_{L} g^{1/2}\lambda^{3/2}/\mu_{L}$, the ratio of 
density, $\rho_{G}/\rho_{L}$, the ratio of viscosity, $\mu_{G}/\mu_{L}$ and
 the Bond number, $B=\rho_{L} g \lambda^2/\sigma$, where the subscripts 
${\small L}$ and ${\small G}$
denote respectively the liquid and gas phases.  Equations~(\ref{N-S}) in 
dimensionless form then read
\be
\p_t {\bf u} = - {\bf u} \cdot \nabla\, {\bf u} + \frac{1}{\rho}
\left[-\nabla p + \frac{1}{Re} \nabla \cdot (\mu {\bf S}) + \frac{1}{B}
\kappa\delta_S{\bf n}\right] + {\bf g_u}, \label{N-S2}
\ee
where ${\bf g_u}=(0,-1)$ represents a unit gravitational force. The 
dimensionless density $\rho$ and viscosity $\mu$ in~(\ref{N-S2}) are 
expressed in each cell, in the term of the volume fraction $C$ as follows:
\be
\rho =  C + (1 - C)\, \rho_{G}/\rho_{L} ; \, \,
\mu = C + (1 - C)\, \mu_{G}/ \mu_{L} .
\ee

Instead of calculating the curvature of the interface, the (dimensionless)
capillary force, $\kappa \delta_S{\bf n}/B$, is represented as 
the divergence of the capillary pressure tensor ${\bf T}$ which is defined
as
 $({\bf I} - {\bf n} \otimes {\bf n}) \,\delta_S/B$,
where ${\bf I}$ is the unit tensor $\delta_{ij}$. In this way, the capillary
force is taken into account as a stress tensor which is computed from 
the gradient of the volume fraction $C$. Briefly, the Navier-Stokes 
equations~(\ref{N-S2}) are solved using finite differences on a staggered
({\small MAC}) Eulerian grid and split-explicit time differencing scheme, 
while incompressibility~(\ref{incom}) is enforced using an iterative 
multigrid Poisson solver.  For full details of the method, readers are
referred to Lafaurie {\it et al.}${}^{15}$ and Li,${}^{16}$ in which 
various tests of accuracy and validity are also given.

Previous numerical studies demonstrated that the details of wave breaking
are very sensitive to initial conditions. Usually, two kinds of initial
conditions have been used to generate wave breaking, one is the Stokes wave
with a pressure forcing at the surface,${}^{2,4}$ another corresponds 
to an unsteady wave with a large enough amplitude.${}^{3-5}$ 
We present here the results for 
waves developing from the initial condition that corresponds to a Stokes
wave in infinite depth calculated at the third order of wave amplitude $a$.
Let $x$ and $y$ denote respectively the horizontal and vertical coordinates, 
and the origin be taken in the center of the cavity, then using the present
scaling, the wave profile, $\eta$, of slope $\epsilon$ 
($=2\pi a/\lambda$) is as follows
\be
\eta = \frac{1}{2\pi} [(\epsilon + \frac{1}{8} \epsilon^3) \cos(2\pi x)
 + \frac{1}{2} \epsilon^2 \cos(4\pi x) + \frac{3}{8} \epsilon^3 \cos(6\pi x)].
\label{profile}
\ee

The numerical results of $Re=2 \times 10^4$, $\rho_{G}/\rho_{L}=10^{-2}$, 
$\mu_{G}/\mu_{L} = 0.15$ and $B=10^4$ are presented in figures 1-3.
The initial wave slope $\epsilon$ is set to be 0.55. Since it is not a
steadily travelling wave and is steeper than any irrotational steady wave,
such initial conditions evolve to breaking as shown below.
 Computation was done on a $256 \times 256$ uniform mesh with a constant
 time step of $10^{-4}$.

Figure 1 shows a time evolution of liquid-gas interface and vorticity
field up to time $t=6$ (we recall that the present length and  time scales 
were chosen such that a linear cosine wave has a period of $(2\pi)^{1/2}$). 
(Complete movie stored as {\small MPEG}
is available on the {\small LMM WWW} server
 ({\tt http://www.lmm.jussieu.fr/Animations.html}).
The initial wave given by~(\ref{profile}) has its crest at $x=0$
and progresses from left to right.  At $t=0$, the velocity field in
 the liquid part is obtained from the velocity potential at the
 third order (without surface tension, i.e. $B\rightarrow \infty$), 
while  the gas is at rest, and then motion is
generated in the bulk of the gas phase 
through dynamical coupling with liquid at the
 interface. 
As a periodic boundary condition is imposed in the direction of wave
propagation, the fluid moving out of the domain on the
 right will rejoin it on the left.
As can be seen, the wave breaks in 
the form of plunging.
To distinguish the two phases, a color jump is used with green
representing the liquid phase and grey the gas phase. However, in each phase,
the values of vorticity are represented by different color levels; the 
maximum vorticity is shown by red and the minimum by blue. 
As shown in Fig.1(a), during the pre-breaking stage, the wave profile, 
particularly near the crest, becomes more and more asymmetric. The wave
 breaks at the time where the whole front face of 
the crest steepens and becomes vertical, and then a jet of liquid is projected
forward into a characteristic overturning motion (Fig.1(b)). The vorticity 
in the liquid phase is mainly located in the vicinity of the tip of the jet
 where the curvature of the interface is important. Under the influence
of gravity, the jet plunges down into the surface below, entraining 
a volume of gas (Fig.1(c)).
The main effect of the impact of the jet is the generation of vortical 
motions under liquid and the formation of splash-up as shown in Fig.1(d). 
Strong vorticity is also generated at the crest
of the secondary jet due to the splash-up. As pushed forward by the
 plunging jet, the second jet is growing in size (Fig.1(e)) and
 plunging (Fig.1(f)), 
generating second splash-up (Fig.1(g)). Third and fourth splash-ups
are observed at $t=3.48$ and $t=5.12$, respectively. 
As shown in Fig.1(h), the wave loses most of its potential 
energy at $t=6$, and splashing no longer occurs. 

In order to gain further insight into the vorticity arising from the 
wave breaking, a close-up view of the active region in Fig.1(d), 
where the first splash-up is formed by plunging jet, is shown in figure 2.
The velocity distribution of Fig.2(a) shows that the flow near the 
plunging point displays a  structure composed of two large vortices, 
one revolving clockwise (beneath the jet) and the other
anticlockwise (in gas). The present mode of splash-up corresponds to
one of modes described by Peregrine${}^6$ (see his Fig.4(b)): i.e. the jet
penetrates the surface, and then, because of its forward motion and 
downward momentum, it acts like a solid surface and ``pushes up'' a jet
of previously undisturbed flow, causing the splash-up. The vorticity
 distribution in this region is shown in Fig.2(b). The positive vorticity
levels (solid line) are mainly concentrated in the regions where the
first splash-up was formed, and at the crest of the second jet, while 
the negative vorticities (dashed line) are mainly located in
 the vicinity of entrapped gas.
The change of direction in velocity and of sign in vorticity near
the plunging point indicates that there exists a strong shear layer there.

Our numerical results of
successive splash-up cycles and vortex-like motions agree  qualitatively
with the laboratory observations of Bonmarin.${}^1$ However, the resolution
($256\times256$) appears to be not fine enough; some of fragments of the 
liquid, as displayed in figures 1-2, does exist, which are due to
the discretisation rather than physical effects.
Simulations of higher resolution on grids of 
$512 \times 512$ and $1024 \times 1024$ 
are in progress; more detailed results will be reported in the near future.

Let $E_k$, $E_g$ and $E_{\sigma}$ denote respectively the kinetic
 energy, the gravitational potential energy and the capillary energy,
 calculated in the liquid part ($C \not= 0$) over one wavelength:
\be
E_k = \frac{1}{2}\int\int_{C \not=0} \rho {\bf u}^{2} dxdy; 
E_g = \int\int_{C \not=0} \rho y dxdy; 
E_{\sigma} = \frac{1}{B} (\Gamma - 1),
\ee
where $\Gamma$ is the total arclength of liquid-gas interface. The total
(mechanical) energy, $E$, is then obtained by summing these three parts.

The time evolutions of normalized values (by its initial
value) of $E$, $E_k$ and $E_g$, are plotted in figure 3. The initial
values of the energies are respectively $E_k = 2.66 \times 10^{-3}$,
$E_g=2.23 \times 10^{-3}$ and $E_{\sigma}=1.13 \times 10^{-5}$.
This figure displays
distinctly different regimes of the energy variation during the
wave breaking process. Before the formation of the jet, the  
kinetic and gravitational potential energies decrease smoothly.
After the jet formation ($t>t_1$) and under the influence of gravity,
the jet plunges down, $E_k$ increases up to the time of impact of the jet
onto the forward surface. The curves of $E_k$ and $E_g$ exhibit some small
oscillations due to the formation of successive splash-up cycles
generated by the first falling jet. Energy is dissipated by viscosity,
therefore the total energy $E$ decreases with the time.
Its evolution, however, is not a linear function of the time.
During the wave evolution, the slope of the total energy curve $E$, up to 
about $t=4$, becomes steeper after the first 2 splash-ups, corresponding to 
a faster energy loss. It can be seen that nearly $70\%$ 
of the total energy of the wave is lost after about three periods.
 
In summary,  an initially steep short gravity wave with surface tension
 undergoing breaking have been simulated numerically; results show good 
resemblance to the laboratory observations of breaking  water wave motions
 including overturning, plunging process, air entrainment, 
and successive splash-up cycles. 
The generation of vorticity both in the liquid and gas by the
 breaking process  has also been  shown. 
Energy dissipation due to viscosity has been found important during
breaking process, particularly at the time of splash-up. We are
currently working on  a faster version of the code in order
to simulate longer wavelengths. 
The ability of the code to simulate the dynamics of the gas
phase also offers the perspective of simulating breaking waves in
presence of wind.

\acknowledgments  The authors wish to thank Prof. D.H.~Peregrine for his
helpful comments and suggesting improvements to the original manuscript,
 and Dr. P.~Bonmarin for discussions. 
The computations were done on the SGI Power Challenge at IRPHE and in part 
on the SP2 of the CNUSC (Centre National Universitaire du Sud de Calcul).
The support from DRET (Direction des Recherches, Etudes et Techniques)
 is gratefully acknowledged.

\newpage

\begin{figure}
\caption{Snapshots of instantaneous liquid-gas interface
and vorticity field of a plunging breaker.
 Displayed region is $-0.5 \le x \le 0.5$ (corresponding to 
one wavelength) and $-0.3 \le y \le 0.3$. Physical parameters and initial
conditions are given in the text.
A color jump is used to simultaneously distinguish the two phases and
represent vorticity: the liquid phase is mostly green 
and the gas phase grey. In both phases, high positive vorticity is red
and very negative vorticity is blue. }
\end{figure}

\begin{figure}
\caption{Close-up view of the active region of figure 1(d). 
The liquid-gas interface is represented by bold line. 
(a) Velocity field. The velocity vectors are plotted
in every two computational cells. Maximum velocity is $0.96$ 
(note that the velocity scale is $(g\lambda)^{1/2})$.  (b) Contours
of constant positive (solid line) and negative (dashed line) vorticity. 
Minimum and incremental levels of vorticity are $-8.91$ and $2.22$.}
\end{figure}

\begin{figure}
\caption{The total energy ($E=E_k + E_g + E_{\sigma}$) (solid line),
the kinetic energy ($E_k$) (dashed line) and the gravitational
 potential energy ($E_g$) (dashdot line), calculated in the liquid phase
over one wavelength, as a function of the time $t$.
Each energy is normalized by its initial value. Marked times correspond to
respectively the formation of jet ($t_1=0.72$), the impact of plunging jet onto
the forward surface and the generation of the first splash-up ($t_2=1.56$), 
the formation of the second splash-up ($t_3=2.84$), the
 formation of third splash-up ($t_4=3.48$) and the formation of fourth 
splash-up ($t_5=5.12$). The total energy $E$ decreases due to viscosity. 
Nearly $70\%$  
of the wave energy is lost after about three periods.}
\end{figure}

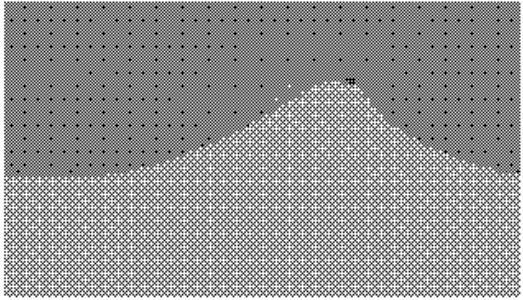
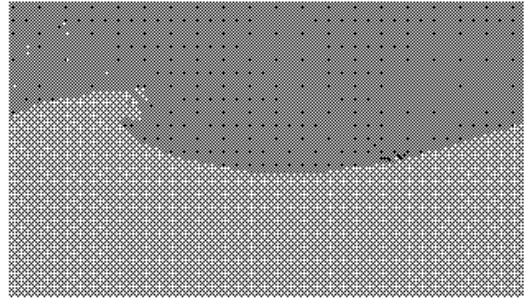
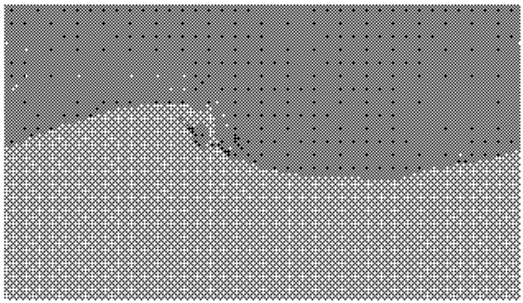
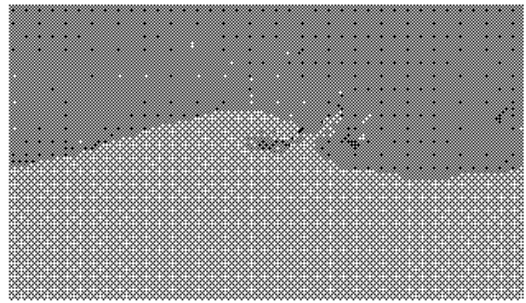
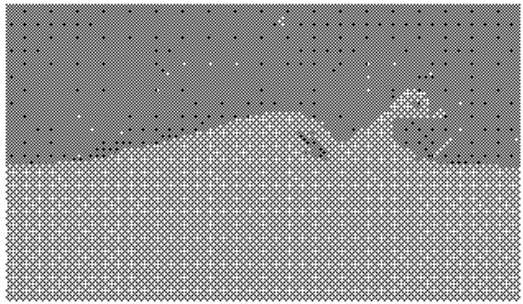
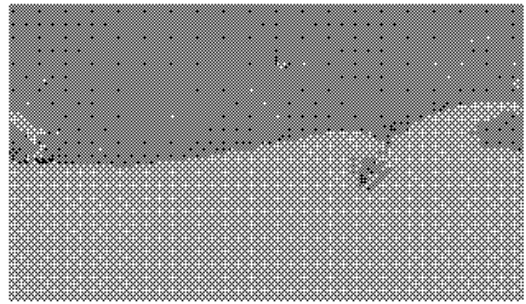
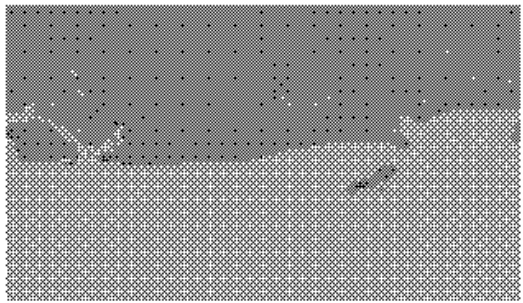
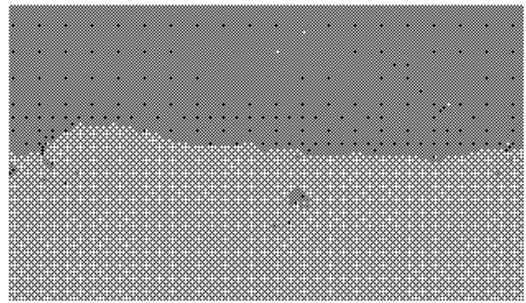
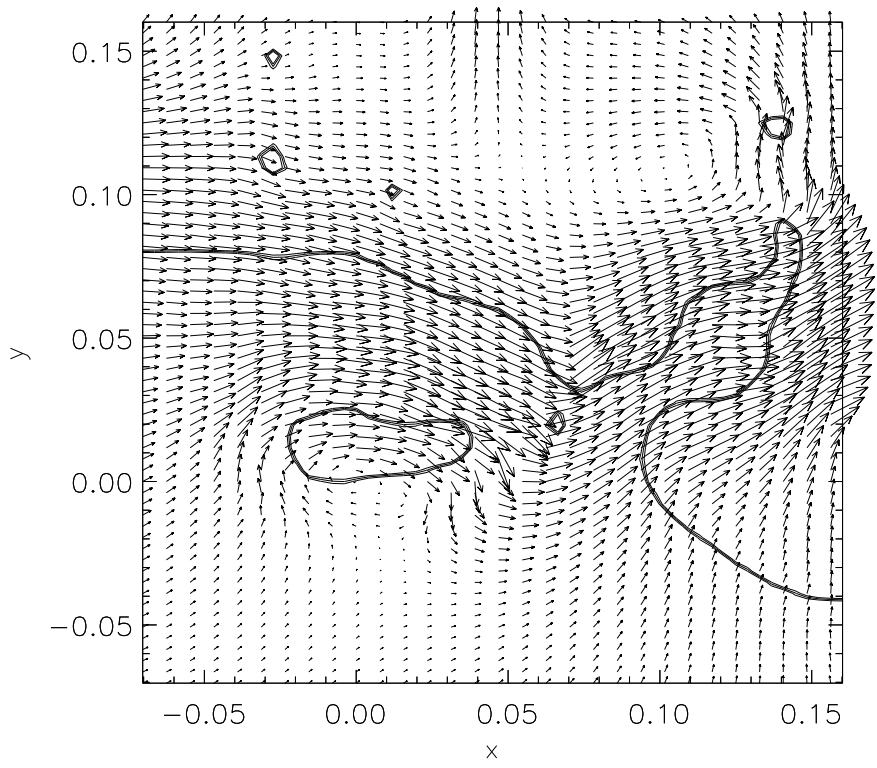
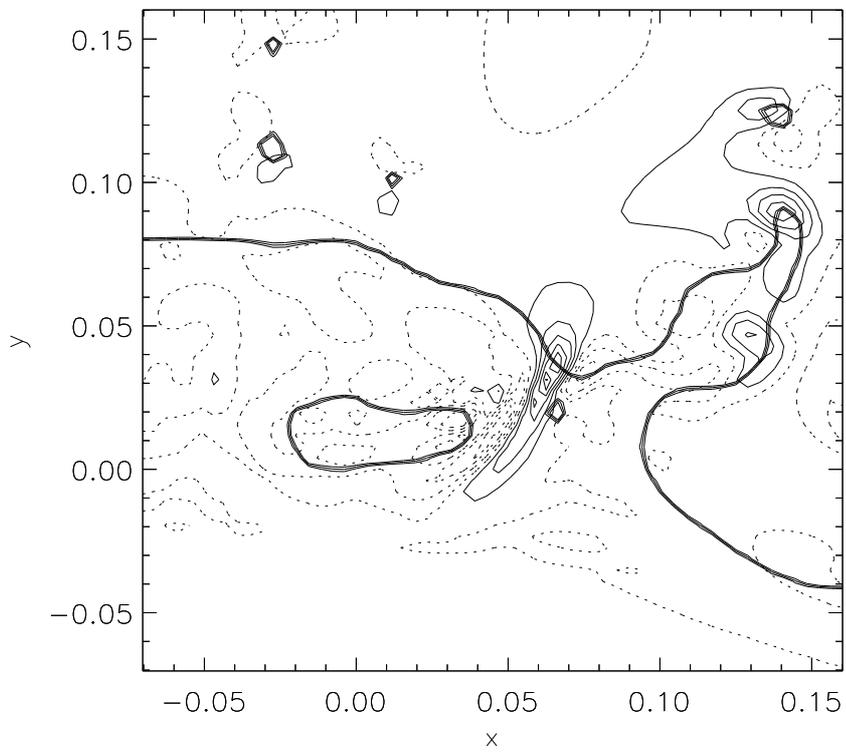
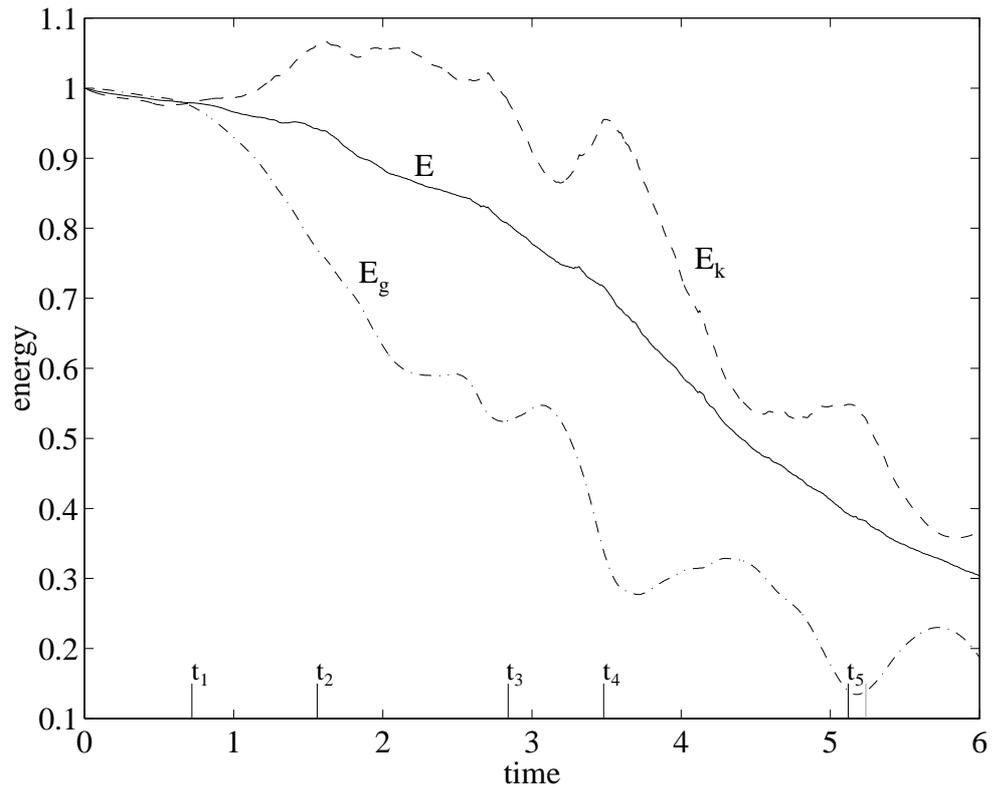
\end{document}